# XPS on corrosion products of ZnCr coated steel: on the reliability of Ar$^+$ ion depth profiling for multi component material analysis

R. Steinberger, J. Duchoslav, M. Arndt, D. Stifter

Christian Doppler Laboratory for Microscopic and Spectroscopic Material Characterization, Center for Surface and Nanoanalytics, Johannes Kepler University Linz, Altenberger Straße 69, 4040 Linz, Austria

Corresponding author: Roland Steinberger, roland.steinberger@jku.at, +43 732 2468 9861;

E-mail: jiri.duchoslav@jku.at, martin.arndt@jku.at, david.stifter@jku.at;

**Abstract**

X-ray photoelectron spectroscopy combined with Ar$^+$ ion etching is a powerful concept to identify different chemical states of compounds in depth profiles, important for obtaining information underneath surfaces or at layer interfaces. The possibility of occurring sputter damage is known but insufficiently investigated for corrosion products of Zn-based steel coatings like ZnCr. Hence, in this work reference materials are studied according to stability against ion sputtering. Indeed some investigated compounds reveal a very unstable chemical nature. On the basis of these findings the reliability of depth profiles of real samples can be rated to avoid misinterpretations of observed chemical species.

## 1. Introduction

In the steel industry new kinds of metallic coatings are steadily in the focus of research and development in order to obtain improved resistance against corrosion. Besides already existing approaches such as hot-dip galvanizing with Zn or ZnMgAl alloys [1] even more sophisticated options are nowadays developed: a very encouraging concept to enhance the protection level involves electro-deposition of ZnCr layers [2], leading also to the need to better understand the formation and structure of corrosion products formed on such novel coatings. One of the methods of choice for such corrosion studies is X-ray photoelectron spectroscopy (XPS) analysis, as shown in [3] for corroded ZnMgAl layers, since this surface sensitive method exhibits a high potential to distinguish between different compounds by resolving their chemical states. In combination with argon ion sputtering XPS depth profiling is usually a suitable technique to investigate critical regions like the layers directly located below the topmost surface or close to interfaces *e.g.* between the steel substrate and the corrosion protection coating [4]. However, it is also known that sputtering by single ions can cause substantial damage of the irradiated material (*e.g.* by preferential sputtering, ion-implantation, atomic mixing, chemical changes, …), leading in the worst case even to a total conversion to a different compound. Consequently, potential modifications of the chemical state of expected corrosion products, especially for novel coating systems like ZnCr, call for a systematic investigation, being the main scope of the current study: for an exact understanding it is mandatory to examine at first the stability of pure reference materials in order to be able to reliably evaluate the composition of real samples with industrial background. In the case of the chosen ZnCr corrosion protection system an insufficient number of studies was conducted to date for the relevant reference compounds, comprising zinc oxide, zinc phosphate, hydrozincite, simonkolleite, Cr (III) oxide, Cr (VI) oxide, Cr (II) chloride, Fe (II) oxide, Fe (III) oxide and Fe (II, III) oxide. Only one detailed study [5] treated the latter mentioned iron oxides. However, the focus was put on single crystals which do not fully meet the expectation how iron is oxidized/corroded in real samples, in contrast to powder samples which should better reflect the relevant corrosion situation. Although Cr (II) chloride is not an expected corrosion product it is also included in our study due to a lack of published data, leading now to first XPS investigations on this material. A further motivation leading to the current work can be found in the circumstance and the observation that even samples possessing a very simple structure can cause severe difficulties in the evaluation of the data obtained by XPS and assisted ion sputtering. Therefore, it is important to direct one´s attention to that set of problems in order to avoid misinterpretations of the acquired data, as might be occasionally observed in some prior studies.

## 2. Materials and methods

### 2.1. Preparation of samples

For our investigations reference materials of high purity were ordered from Sigma Aldrich. The following components were used as delivered: $Zn_5(CO_3)_2(OH)_6$ (hydrozincite), $Zn_3(PO_4)_2$ (zinc phosphate), $Cr_2O_3$ (Cr (III) oxide), $CrO_3$ (Cr (VI) oxide), $CrCl_2$ (Cr (II) chloride), FeO (Fe (II) oxide), $Fe_2O_3$ (Fe (III) oxide), and $Fe_3O_4$ (Fe (II, III) oxide). Additionally, two samples were synthesized by ourselves: ZnO (zinc oxide) by heating hydrozincite for several hours in a laboratory oven at 300 °C and $Zn_5(OH)_8Cl_2 \cdot H_2O$ (simonkolleite) according to a synthesis procedure described [6]. It is known that zinc hydroxide is also a possible corrosion product, but - as reported by Duchoslav *et al.* [7] - $Zn(OH)_2$ is even not stable against X-ray irradiation, hence, an investigation by XPS is not suitable. All materials were used in form of powders and for the analysis most of them were pressed onto pieces of indium foil. The rest of the powder samples which exhibited an insufficient adherence to the foil were filled in a polymer block with a milled circular pocket in the center. It was paid attention to use a hole diameter which is sufficiently large so that the surrounding polymer is not affected by sputtering.

2.2. Instrumentation

All XPS measurements were performed with a Theta Probe system from Thermofisher, UK. For control and data acquisition as well for the data evaluation the Avantage software package provided by the system manufacturer was used. As source of radiation a monochromated Al-$K_\alpha$ X-ray gun (1486.7 eV) was operated at a voltage of 15 kV, with an emission current of 6.7 mA (100 W) and a chosen spot size on the sample of 200 μm. The hemispherical sector analyzer for electron detection was operated in the CAE-mode (constant analyzer energy) with the pass energy for the high resolution scans set to 50 eV, being a reasonable compromise between obtainable energy resolution and intensity. For all measurements the step width of the energy channels was taken to be 0.1 eV. For charge neutralization on the sample surface a standard dual flood gun was used, which provides electrons of low energy (typically -2 eV) and in addition a beam of $Ar^+$ ions with low kinetic energy. For sputter experiments the system is equipped with an additional $Ar^+$ ion gun, which was used for this work with a set of standard parameters: an acceleration voltage of 3 kV and an ion current of 1 μA were chosen to provide on one hand a reasonable sputter rate and on the other side to minimize potential surface roughening. The base pressure of the UHV analysis chamber was in the low $10^{-9}$ mbar range. With the flood gun applied the pressure settles in the lower half of the $10^{-7}$ mbar range. For data analysis the background of the peaks in the spectra is eliminated by applying Shirley background subtraction. The peak-areas were normalized by using Scofield sensitivity factors to obtain the elemental composition of the surface.

2.3. Measurement procedure

In order to avoid cross-contamination by sputtering between different samples, only a maximum number of two were mounted together on a 7 x 7 $cm^2$ sample holder, with substantial spacing in-between them. On each sample a depth profile was performed: in the center of the sputtered crater high resolution scans of the interesting photoelectron levels – and when needed also from Auger transitions – were acquired followed by a subsequent sputter step. Sputtering was started with an initial sputter duration of 400 s carried out 25 times, continued with 10 more steps by sputtering for 2000 s and finally followed by 3 periods each one lasting 4000 s. With the ion gun parameters given in section 2.2 a sputter rate of ~ 3.3 nm/min can be reached on a Si standard. To obtain a reasonable ratio of the area of the sputter crater to the X-ray beam spot size in order to avoid edge effects, the raster area for the crater was set to 2 x 2 $mm^2$.

For the characterization of the iron oxides high resolution scans of the C1s, O1s, Fe2p and FeL$_3$M$_{45}$M$_{45}$ peaks were acquired. For the Cr oxides the C1s, O1s, Cr2p, Cr3p and Cr3s levels were recorded, for CrCl$_2$ the Cl2p peak was additionally included. Zinc oxide and hydrozincite were investigated by acquiring the C1s, O1s, Zn2p and ZnL$_{23}$M$_{45}$M$_{45}$ spectra. Furthermore, zinc phosphate and simonkolleite required acquisition of the P2p and Cl2p peaks. A charge shift correction according to adventitious carbon was not necessary for this study due to the fact that absolute binding energy values were of no concern for data evaluation, and in some cases were even not possible because of a lack of adventitious carbon on the surface. Instead, the concept of the modified Auger parameter, which is a powerful method to resolve and identify chemical states, was additionally applied for data analysis [8].

## 3. Results and discussion

In the following sections a systematic description of the behavior during sputter depth profiling of different compounds, relevant for corrosion formation on ZnCr coatings, is presented and discussed.

3.1. Zinc oxide and zinc phosphate

A compound which is frequently formed when Zn and Zn-based protection layers are exposed to a corrosive environment is ZnO. In Fig. 1 the Zn2p and O1s spectra during long time sputtering of this compound are presented as first part of the results. One can observe that with increasing sputter time no considerable changes, according to the peak shape, occur; only the intensities of the Zn2p and O1s signals are lower at the surface level (before sputtering) since some adventitious carbon (~ 18 at.%) was initially present on the topmost surface. The small shift of the oxygen peaks, observed between the original surface and the subsequent levels, occurs in all other measured spectra exactly in the same way and can be related to slightly different charge neutralization conditions on the surface. The ratio of zinc to oxygen is stable with 55 at.% to 45 at.%. From the theoretical stoichiometry one would expected a ratio of one to one, but prior experience has shown that the spectra and measured peak ratio are related to zinc oxide. In addition the modified Auger parameter was calculated as a function of the sputter time, resulting in a quite constant value of 2010.06 eV ± 0.04 eV, which is indicative for ZnO [9,10]. The results are apparently indicating that sputtering – also applied for long time – does not significantly modify zinc oxide.

A further possible corrosion product identified is zinc phosphate, however, with the formation being induced by the presence of phosphates. In Fig. 2 the spectra of the Zn3s, O1s and P2p photo peaks are shown as a function of the sputter time. For the sake of convenience the Zn3s peak is presented instead of the Zn2p peak, since the Zn3s level is situated within the scanning range of the P2p photo peak. The differences in intensity between the surface level and the other levels of the different scans are caused by a low contamination of adventitious carbon on the surface, which vanished immediately after the first sputter cycle. Again, the peak shifts between the surface levels and the deeper levels are very similar for all spectra and are, therefore, not related to a change in chemistry. It is equally observed that with increasing sputter duration no remarkable changes occur. The theoretically calculated stoichiometry for this compound is 15.5 at.% of phosphor, 23 at.% of zinc and 61.5 at.% of oxygen, which reflects the experimental result of 15 at.% P, 28 at.% Zn and 57 at.% O. The concentration exhibits a good stability over the whole experimental time. In addition, to confirm the chemical stability during sputtering, the modified Auger parameter was calculated for each level and found to be constant around 2009.2 eV. If all these observations are considered, it can be concluded that XPS depth profiles performed on zinc phosphate will deliver correct data.

3.2. Hydrozincite and simonkolleite

A corrosion product that is frequently present on Zn-based coatings is hydrozincite, which can be formed in an atmosphere containing $CO_2$. In Fig. 3 the O1s peak is plotted and the elemental composition as function of the sputter time is shown, with the Zn2p peak given in the inset. The large shift of about 1.7 eV of the O1s peak at the surface level immediately attracts attention. In contrast, the shift of the Zn2p peak is approximately only 0.5 eV. The carbonate peak is even shifted into the opposite direction by 0.1 eV to 0.2 eV, indicating that this behavior is not related anymore to charging but is induced by a varying chemical shift. In addition, the modified Auger parameter averaged over all levels gives 2009.83 eV ± 0.12 eV, with the standard deviation being slightly increased. In addition, the mean value does not match the empirical value of hydrozincite, which is settled around 2009.2 eV to 2009.3 eV. If the modified Auger parameter is considered level by level, following observation can be made: at the surface the value corresponds well with hydrozincite but with rising sputter time the parameter approaches 2009.9 eV, pointing to the formation of ZnO. Also the feature on the left side of the O1s peak is an indicator and the peak distance Zn2p to O1s with 491.22 eV additionally matches that one of ZnO, which is 491.25 eV. The concentration plot of zinc, oxygen and the carbon attributed to carbonate (Fig. 3) reveals a fast decrease of the carbonate content accompanied by a reduction of the oxygen content and a dramatically increased zinc occurrence. At higher sputter time the ratio of Zn to O is very similar to that found for the measurement of ZnO. It seems that the carbonate content is affected by preferential sputtering. This observation is supported by regarding the different sputter yields for the different elements which were determined by using the TRIM (Transport of Ions in Matter) software [11]. Of course it is not possible to directly relate the changes in composition observable in the spectra to the calculated yields - which are just given here as reference points - since the real sputter process of such compounds is generally too complex for being fully described by such a simulation. However, the resulting sputter yields (given in sputtered atoms per $Ar^+$ ion) are 2.64 Zn atoms/ion, 0.24 C atoms/ion and 4.2 O atoms/ion. If we assume that the real sputter yield of the carbon atoms is much higher due to their bonds to the preferentially sputtered oxygen atoms, a conversion from hydrozincite to a compound containing as main component zinc oxide is very likely. Nevertheless, the presence of a small amount of remaining hydrozincite with depleted carbonate content cannot be excluded. As additional remark it has to be mentioned that changes corresponding to the chemical stoichiometry may not only be caused by sputtering, but a slight and steady decomposition trend of hydrozincite was also observed during standard XPS measurements when long acquisition times were necessary, possibly caused by the X-ray radiation under the same conditions as already reported in ref. [7].

If a Zn or ZnCr coated material is exposed to an environment which contains chlorine, *e.g.* in form of NaCl, the formation of simonkolleite is possible for certain pH values. The sputter results of this compound, presented in Fig. 4, bear a resemblance to hydrozincite. Especially the shift of the oxygen surface level of about 1 eV compared to the chlorine peak (~ 0.9 eV, not shown in Fig. 4) and the zinc peak (~ 0.1 eV), which are shifted in the opposite direction, indicates a chemical change. Again, the mean value of the modified Auger parameters was determined (according to our experience it should be similar to hydrozincite): 2009.75 eV ± 0.08 eV. In detail, the value at the surface level was found to be 2009.4 eV and at the end of the sputter experiment slightly more than 2009.8 eV. In addition, except for the surface level, a shoulder can be observed on the left side of the O1s peak. All these observations hint at a formation of ZnO, together with the calculated distance between the Zn2p and O1s peaks (491.21 eV), positively contributing to this assumption. The elemental distribution over sputter time, which contains the spectra of Zn2p as function of time, is also shown in Fig. 4. Initially, the stoichiometry is close but not exactly meeting the expectations with slight differences to the structural formula. However, when sputtering has started remarkable changes take place: the Cl content is rapidly decreased by one third, and further slightly decreasing, the oxygen level also drops from around 50 at.% to 40 at.% and consequently the relative zinc concentration is increased. If all this knowledge is taken into account it indicates that simonkolleite is degraded due to sputtering, with one possibly formed main product being ZnO. However, the observed stoichiometry and modified Auger parameter, being more or less a linear combination of the modified Auger parameters of all present compounds, hint at a possible presence of at least one additional Zn based compound; if it is remaining simonkolleite with lowered chlorine content or a new compound cannot be resolved.

### 3.3. Chromium (III) oxide, chromium (VI) oxide and chromium (II) chloride

Especially if ZnCr coatings are used to protect steel parts a likely corrosion product is chromium (III) oxide, but also Cr (VI) may be found under certain conditions. In Fig. 5, the spectra of a Cr (III) sample are shown. The sharp structure directly located on the right side of the $Cr2p_{3/2}$ peak at the surface level is related to a multiplet splitting due to the fact that $Cr_2O_3$ has unpaired electrons ([Ar] $3d^3$) [12], also causing the asymmetric shape on the left side of the peak. In the spectra of Cr3s also a multiplet splitting can be observed. Since only Cr (III) - out of the possible oxidation states of Cr - shows a really pronounced multiplet splitting, these features are a reliable proof for the presence of the correct reference material. During the whole time of sputtering all spectra show a stable behavior, only worth mentioning is that the $Cr2p_{3/2}$ surface level shows the double peak caused by multiplet splitting. However, the asymmetry of that peak, also induced by multiplet splitting effects, remains observable during sputtering. The elemental composition does not fully meet the stoichiometric value of 40 at.% Cr and 60 at.% O, since the experiment reveals approximately 33 at.% Cr and 67 at.% O, with the oxygen signal slightly decreasing with time (about $5 \cdot 10^{-5}$ at.%/s). The latter observation is most probably related to preferential sputtering of oxygen since the sputtering yield of oxygen is 5.8 atoms/ion compared to that one of chromium being only 2.5 atoms/ion [11]. To be sure that these results are reliable and not influenced by *e.g.* preparation artifacts, the experiment was repeated with Cr (III) oxide obtained by thermal treatment (in a standard laboratory oven at 200 °C for 7 h) of $(NH_4)_2Cr_2O_7$ which is a Cr (VI) compound. It turned out that the results could be well reproduced, including also a recognizable multiplet splitting. Afterwards, for all experiments the sputter crater was inspected and it was observed that the material turned from green to brown, meaning that the physical properties of that material have changed. However, Cr (III) oxide can be sputtered, even for a long time, and still be afterwards recognizable as Cr (III) oxide in the XPS spectra.

In contrast to Cr (III) oxide, Cr (VI) oxide is strongly toxic. Due to that fact, industry is bound by strict regulations for the usage of this material. Hence, it is of high interest to obtain information on the behavior and the chance to detect Cr (VI) oxide during investigation by XPS. In Fig. 6 the spectra of Cr2p and Cr3s are shown as an example out of all resulting spectra. The most striking feature can be found on the surface level of the $Cr2p_{3/2}$ photo peak, which is not related to multiplet splitting because the electron configuration of Cr (VI) belongs to that one of a diamagnetic state. The evaluation of all spectra finally indicated that some Cr (III) oxide is additionally present next to the original Cr (VI) oxide, which could be confirmed by a comparison to the reference spectra of $Cr_2O_3$ shown above. After one step of sputtering the peak shapes are totally changed and remain then stable during further ion bombardment: the structures on the left sides of the $Cr2p_{1/2}$ and $Cr2p_{3/2}$ vanish and also the Cr3s spectra reveal the change of the chemical properties. At the surface level the Cr (VI) peak overlapps with the multiplet splitting feature of Cr (III) at higher binding energies, resulting in a two peak structure where the left one is substantially higher. Since it is known that the separations in multiplet splitting in the 3s photoelectron level of first series of transition metals is characteristic for the chemical state [13], this approach can also be taken for the identification of different compounds. In this case a separation of 4.00 eV ± 0.08 eV is obtained, averaged over all levels not taking the surface into account, which is in good agreement to that one of the original Cr (III) oxide with 4.07 eV ± 0.08 eV and to 4.1 eV found in another work [14]. The further progression of the elemental composition in dependency of the sputter time shows a behavior close to that already observed for Cr (III) as shown above. In detail, the composition starts at a Cr to O ratio of about 31 at.% to 69 at.% (without taking the surface level into account) and the oxygen signal slowly decreases in time (about $3 \cdot 10^{-5}$ at.%/s), related to preferential sputtering. Once more the material $(NH_4)_2Cr_2O_7$ which corresponds to Cr (VI) was additionally investigated to confirm these results. The shape of the surface level peaks of Cr was similar, also indicating that a small amount of Cr (III) is already present at the beginning. After one sputter cycle no Cr (VI) oxide could be detected any more and in addition the stoichiometry matches that one as determined before. Summarizing, it was found that even a short sputter time already causes a dramatic change of the chemical properties of these compounds, *i.e.* a rapid conversion of Cr (VI) to Cr (III), making XPS depth profiling unsuitable for the detection of potentially embedded Cr (VI) contents.

Based on that findings it is also interesting to investigate a further oxidation state of Cr, *i.e.* $CrCl_2$ was taken with chromium being present as Cr (II), although this material is not a usual corrosion product since it is known to be of very hygroscopic nature and furthermore not stable in a corrosive atmosphere. On one hand Cr (II) easily tends to be oxidized in air to higher oxidation states and on the other hand it could be reduced to a lower state by the sputtering process. Due to the known instability at ambient conditions the sample preparation and mounting was carried out as fast as possible. Nevertheless, small regions, especially next to the borders of the powder patches turned to green (from an original grey color), which hints to an oxidation process, *e.g.* leading to the formation of the dark green chromium trichloride hexahydrate ($CrCl_3 \cdot 6H_2O$). Therefore, Fig. 7 shows the Cr2p peak from the center region of the material with the behavior of the elemental concentrations over sputter time, and with the spectra of oxygen given in the inset. The substantial amount of oxygen at the surface level clearly shows two kinds of oxygen, with one most likely related to crystal water and the other one to the exposure of the material to air during sample preparation. Consequently, a part of the chromium is also present in oxidic form but the amount of oxygen becomes depleted in time due to a limited depth of diffusion or due to preferential sputtering being also an influencing factor, since the sputter yield of O in Cr (III) oxide is much higher (2.3 times) than that of Cr. In detail, by observing the progress of the O1s signal over time it seems to be that the crystal water is rapidly removed by sputtering. Also for the Cr2p spectra a significant change over time is evident. Taking properties like FWHM, multiplet splitting in the Cr3s spectra (not shown here, but also a multiplet splitting is roughly visible at the Cr2p$_{3/2}$ surface level), ratio of the concentration chromium to oxygen, peak distances between different elements, etc. into account, we can conclude that within the first levels the main peak of Cr2p originates from Cr (III) oxide and chloride. The peak which rises at lower binding energy is attributed to Cr (II) chloride due to the fact that metallic Cr does not reveal multiplet splitting and such an effect is visible for all times in the corresponding recorded Cr3s spectra. Also the peak separation between a metallic and oxidic peak should be about 0.9 eV higher as shown in former measurements concerning samples containing metallic Cr and Cr (III) oxide. Fitting of the Cr peak into one part which belongs to oxidation state III and another one attributed to Cr (II) chloride leads to the estimation of an average elemental concentration hinting at $CrCl_2$ (Cr 39 at.% and Cl 61 at.%, ± 4.0 at.%). Preferential sputtering, with 1.02 Cr atoms/ion and 3.83 Cl atoms/ion [11], explains the larger error range in the composition. In one work [14] the multiplet splitting of $CrCl_3$ is stated for Cr3s to be 3.8 eV; in our experiments, fitting of the 3s level with a Cr (III) and a Cr(II) component leads to a similar value for the multiplet splitting of the Cr (III) contribution. An appropriate reference for the noticeable multiplet splitting of the paramagnetic Cr (II), as indicated by our data, could not be found in the literature.

3.4. Iron (II) oxide, iron (III) oxide and iron (II, III) oxide

If the corrosion process is well advanced beyond the protective range of the coating then also metallic iron at the interface - steel substrate to protective coating - starts to oxidize and form corrosion products, which are also necessary to be taken into account. Fe (II), Fe (III) and also Fe (II, III) oxides were investigated with the spectra of the Fe2p photoelectron level shown in Fig. 8, where the 2p$_{1/2}$ and 2p$_{3/2}$ peaks as well the corresponding shake-up satellites can be observed.

Starting with Fe (II) oxide, the shift between the surface level and the other levels (*e.g.* for oxygen) cannot be just explained by charging. Here it is related to a change of the chemical state, since the O1s peak is just slightly shifted (~ 0.1 eV) whereas the Fe2p$_{3/2}$ level much more (~ 1.4 eV). Similar to multiplet splitting also the separation of the shake-up satellites to their photoelectron lines carry unique information about the oxidation state [13]. By comparing the distances of the photo peaks to the corresponding shake-up satellites (surface ~ 7.9 eV, rest ~ 6.1 eV) it is revealed that the surface initially consists of Fe (III) oxide and the other levels represent Fe (II). For cross checking, nearly identical separations could be deduced from the work of Chuang *et al.* [5]. The circumstance that the surface of a Fe (II) sample, in course of this study a powder material, might be covered with a thin Fe (III) layer can also be found in ref. [15]. In addition the surface composition of Fe to O was found to be around 23 at.% to 77 at.% which also indicates the presence of a compound different than Fe (II). Investigating the further levels brings up that peak positions, shapes and shake-up satellites show no peculiarities despite the fact that on the right shoulder of the Fe2p photo peaks, as well for spin-up and also spin-down state, a peak structure grows with increase in sputter time which is related to a reduction of the oxidic iron to metall. The stoichiometry of the first few sputter cycles is related to FeO since the composition is about 43 at.% Fe and 57 at.% O and it is known that this kind of iron oxide is in reality not totally following the theoretical stoichiometry. However, more of concern is the rising of the metallic part causing an increase of the signal of iron in time related to the reduction of Fe (II) oxide.

Continuing with the spectra of Fe (III) oxide, again from the surface to deeper levels a change in chemistry of the sample material can be found. Observing the distance between Fe2p$_{3/2}$ peak and the corresponding shake-up satellite results in a separation of 8.1 eV for the surface and of 6.0 eV averaged over all other levels, which supports the results of the Fe (II) measurement that there Fe (III) is located on top. Unfortunately, the determined surface concentration of 30 at.% Fe and 70 at.% O deviates from the expected theoretical stoichiometry of 40:60. The further trend shows a steady decrease of oxygen and increase of iron. On the basis of all this information, especially taken from the separations of the shake-up satellites, it can be concluded that already after the first time of sputtering Fe (II) is the most dominating compound in the Fe2p spectrum. Just a few sputter cycles later the appearance of iron in metallic form is obvious and increasing in time. Again, the sputter yield in the Fe (III) compound is 2.45 Fe atoms/ion and 5.87 O atoms/ion. For Fe (II) oxide 3.07 Fe atoms/ion and 4.92 O atoms/ion were determined by using the TRIM software, with all values indicating preferential sputtering of oxygen. Taking all these results and information into account it is evident that iron (III) oxide is reduced to iron (II) oxide and furthermore a reduction to metallic iron takes place, with the formation of the metallic iron being faster for Fe (II) than for Fe (III).

In addition the behavior of Fe (II, III) oxide was investigated during long time ion sputtering. As expected the findings are similar as described in the cases of Fe (II) and Fe (III) above, meaning that the surface consists of Fe (III) with the shake-up satellite being nearly not visible. After one sputter cycle Fe (II) oxide is predominant (6.1 eV energy separation between shake-up and main peak) and with increasing sputter time a further rise of the metallic part of iron, which is already appearing after initial sputtering, can be observed. Checking the sputter yields obtained from TRIM simulation give for iron 2.56 atoms/ion and for oxygen 5.41 atoms/ion. Since Fe (II, III) contains bivalent as well as trivalent iron the conversion rate to metallic form is located between that of Fe (II) and Fe (III), meaning that sputtering is again a problematic issue also for this kind of compound.

In general the results obtained from the investigations performed on the different iron oxide compounds are in good accordance to the work described in [5,15], despite the fact that in the present work powder materials, which should correlate more with structures occurring in real corrosion processes, were used instead of single crystals broken under UHV conditions. Only, slightly different rates of forming metallic iron are observed with a faster increase of the metallic part for the powder samples, which is most probably caused by a different sputter rate: *i.e.* in contrast to the described rate of a few Å/min in the other work we estimated on the base of the sputter yields obtained from TRIM simulations a sputter rate of around 7.5 nm/min for all of the used iron oxides.

## 4. Conclusions and outlook

It comes apparent that great care has to be taken while carrying out depth profiling experiments or just a cleaning procedure of the topmost surface of corroded materials when using ion sputtering: selected compounds were proven to be of unstable nature in the context of corrosion studies for novel developed ZnCr coatings. As changes of the chemical properties, most frequently a reduction of the oxidation state of the different reference materials was observed, including Cr (VI) oxide, hydrozincite, simonkolleite and several iron oxides. Fortunately, also more stable ones were found with Cr (III) oxide, zinc oxide and zinc phosphate. In case of the Cr (II) chloride sample, a distinction between the additionally identified Cr (III) oxidation state and Cr (II) was found to be possible. Concerning the sputter profiling itself the particular ion range is estimated to range from 26 Å ($Cr_2O_3$ and FeO) up to 45 Å ($CrO_3$ and simonkolleite) [11]. The escape depth of the hit atoms in the target material is usually below 10 Å [16,17] meaning that next to preferential sputtering also atomic mixing has to be taken into account as possible degradation effect. Further physical processes which may cause the instable behavior induced by ion sputtering and noticeable in *e.g.* the reduction to lower oxidation states are also described elsewhere [5,16,18,19,20]. In addition, it is also necessary to mention that such surface sensitive methods like XPS can be negatively affected by the surface roughness caused by $Ar^+$ ion sputtering. In total, these results should raise the awareness that careful data evaluation and interpretation has to be carried out in case of such depth profiles for identifying the composition of such corroded layers. Even with the above acquired knowledge of the exact chemical evolution of some species it will not always be possible to obtain unambiguous information, so it is mandatory to consult additional analyzing techniques providing complementary knowledge.

Currently, other methods in contrast to etching by single ions are developed, *e.g.* Ar cluster sputtering or buckminsterfullerene ($C_{60}$) ion sputtering. Although these techniques are more complex and cost intensive, it would be essential to study their potential for measurements on corroded matter in the future, since in some other fields promising results could already be achieved [21,22,23,24]. Besides, in order to decrease the damage induced by single $Ar^+$ ion sputtering and to resolve the real composition of a depth profile future investigations could also be aimed at a variation of the sputter parameters (especially acceleration voltage and ion current) or with sample cooling [7] and at performing depth profiles of reference materials which are mixed together, reflecting even better the structure and composition of real corroded material [3]. However, for the actually performed depth profiles the obtained reference sample spectra can serve already as valuable database for such surface studies and to assess the reliability of the corresponding data evaluation.

**Acknowledgements**

Financial support by the Federal Ministry of Economy, Family and Youth and the National Foundation for Research, Technology and Development is gratefully acknowledged.

**Figures**

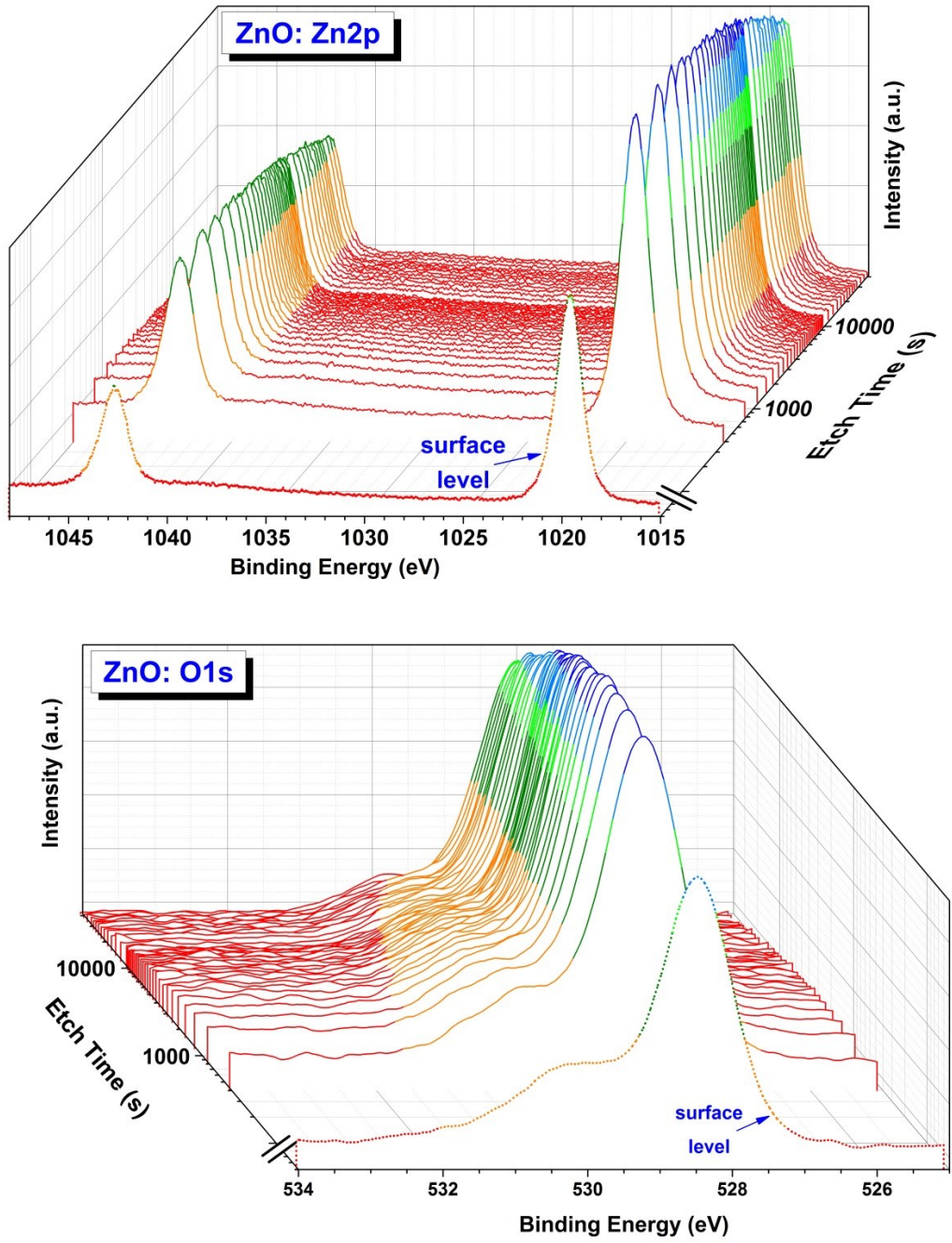

Fig.1. XPS depth profiling of zinc oxide, with high resolution scans of the Zn2p and O1s photo peaks.

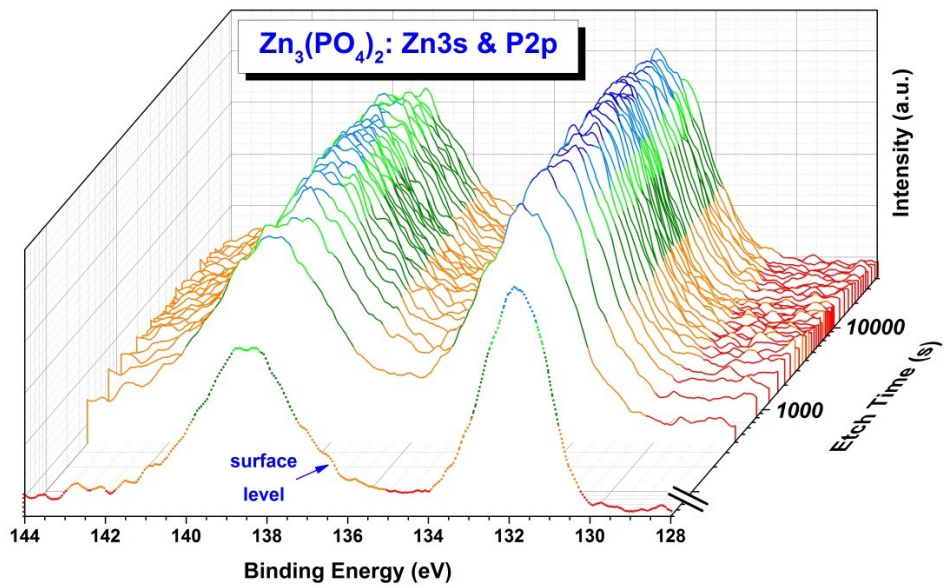
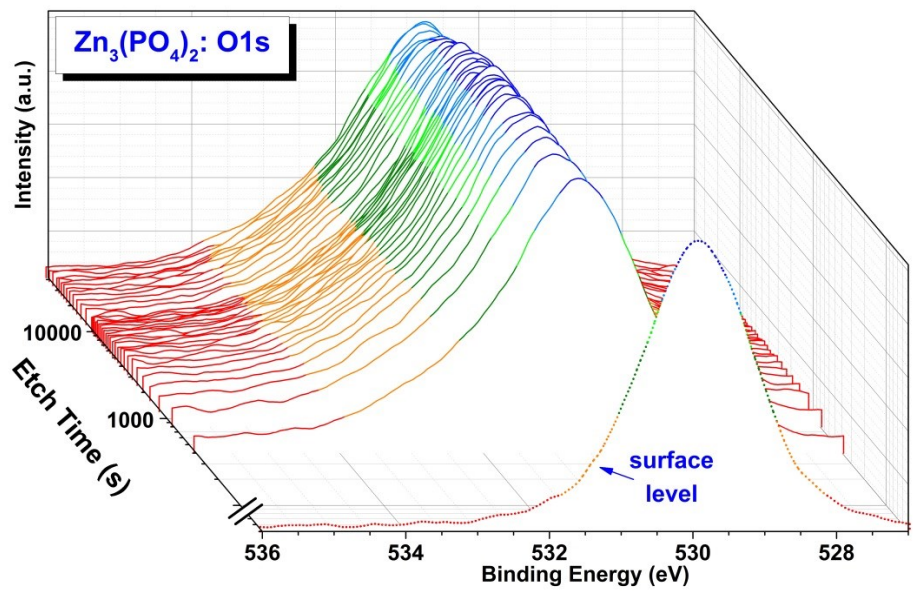

Fig.2. XPS depth profiling of zinc phosphate; high resolution scans of Zn3s, P2p and O1s are shown.

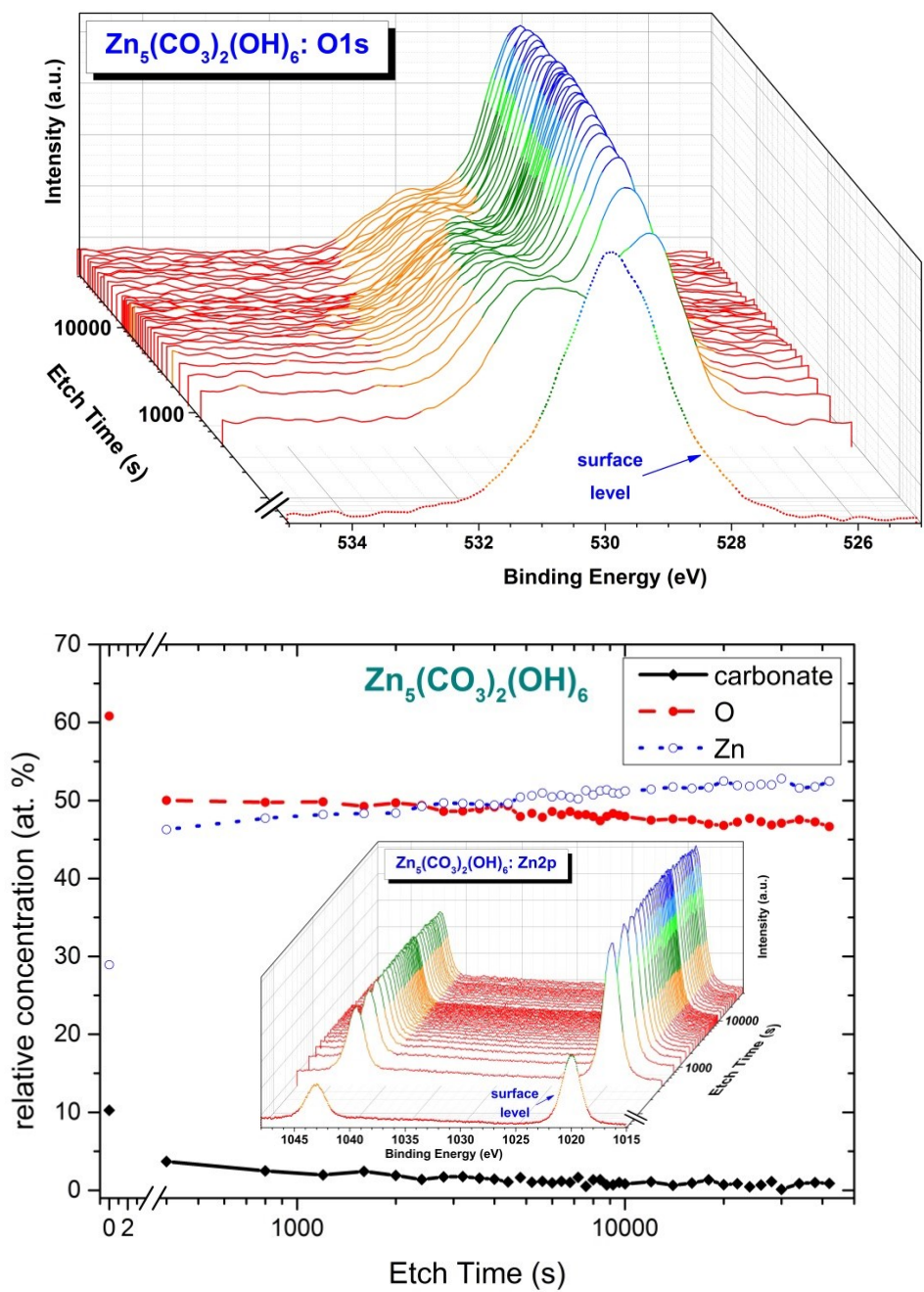

Fig.3. XPS depth profiling of hydrozincite; high resolution scans of O1s and the elemental composition (Zn, O, C attributed to carbonate) over time, including spectra of Zn2p in the inset, are shown.

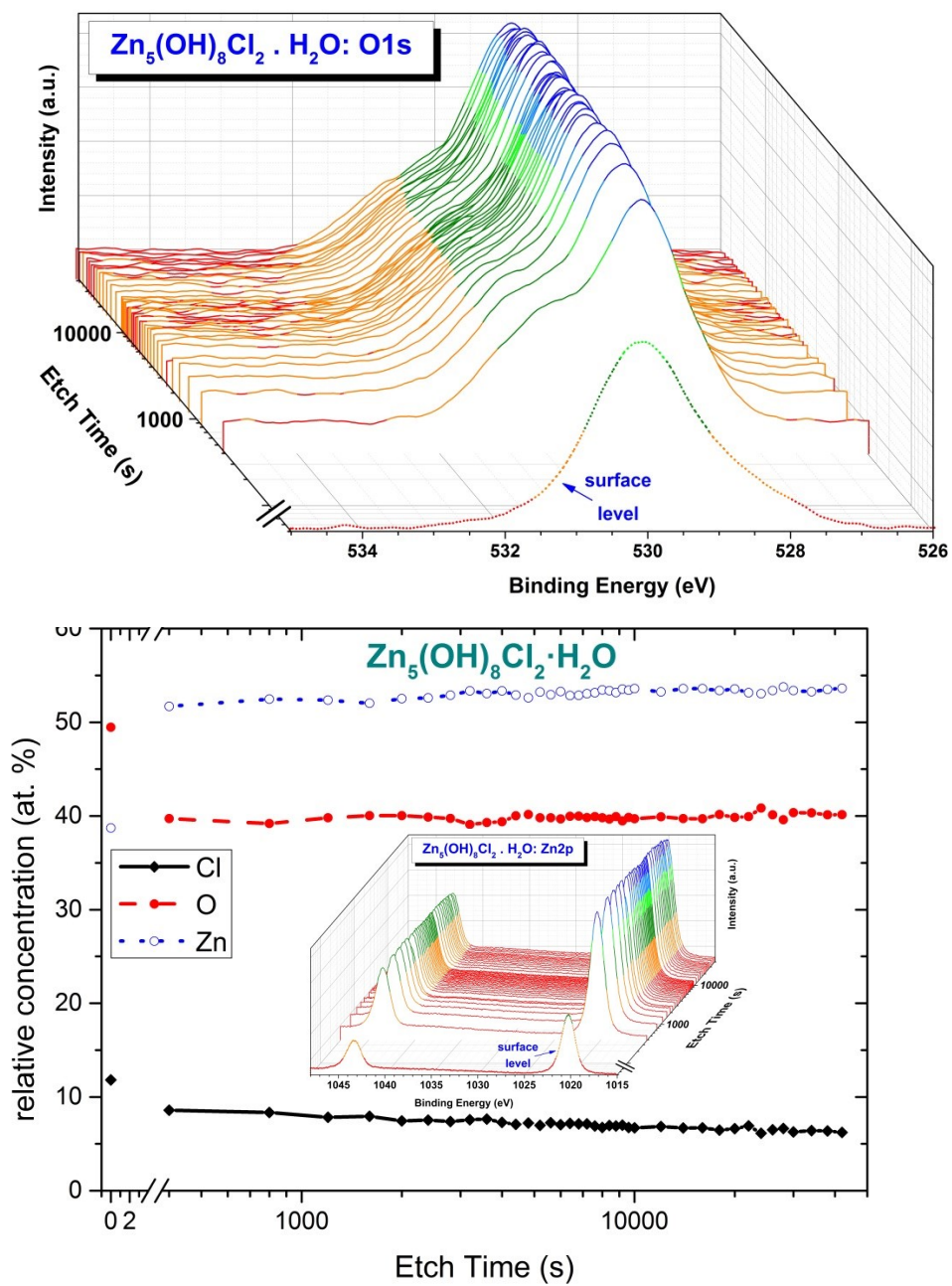

Fig.4. XPS depth profiling of simonkolleite; high resolution scans of O1s and the elemental composition (Zn, O, Cl) over time, including spectra of Zn2p in the inset, are shown.

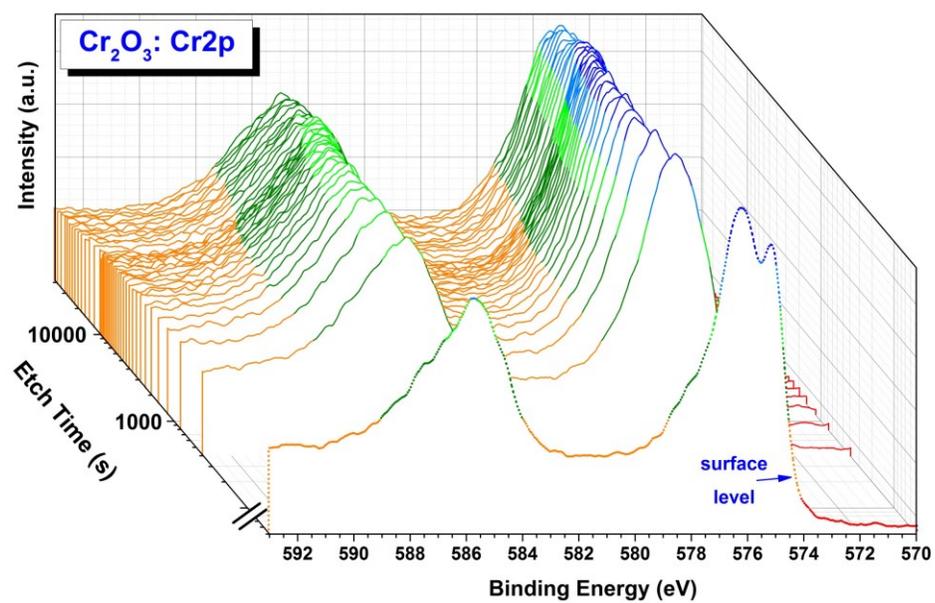

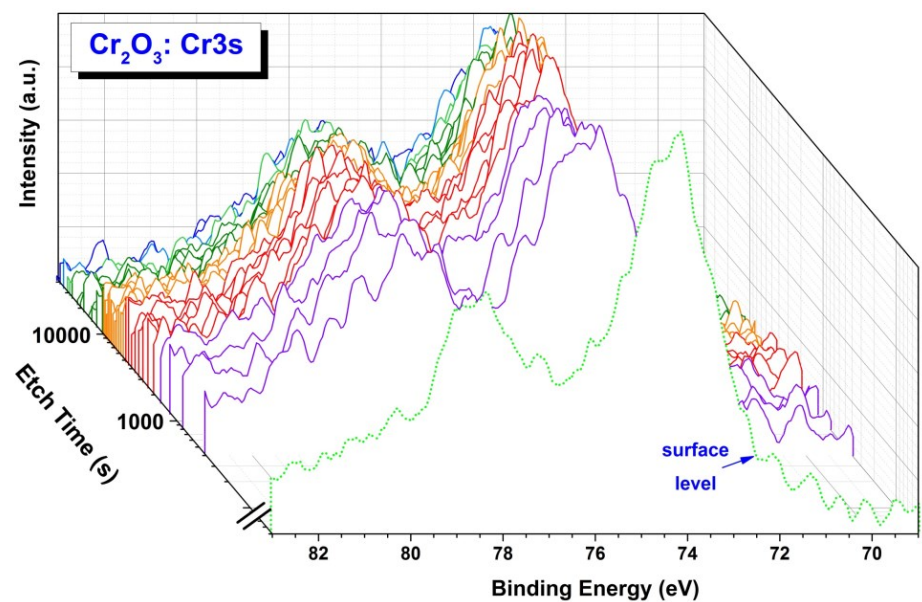

Fig.5. XPS depth profiling of Cr (III) oxide, with high resolution scans of the Cr2p and Cr3s levels.

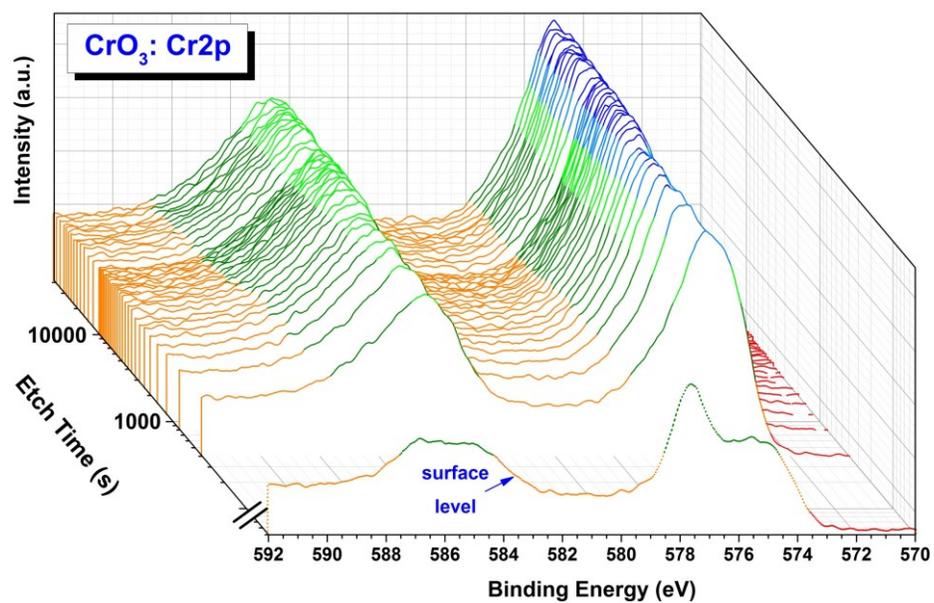
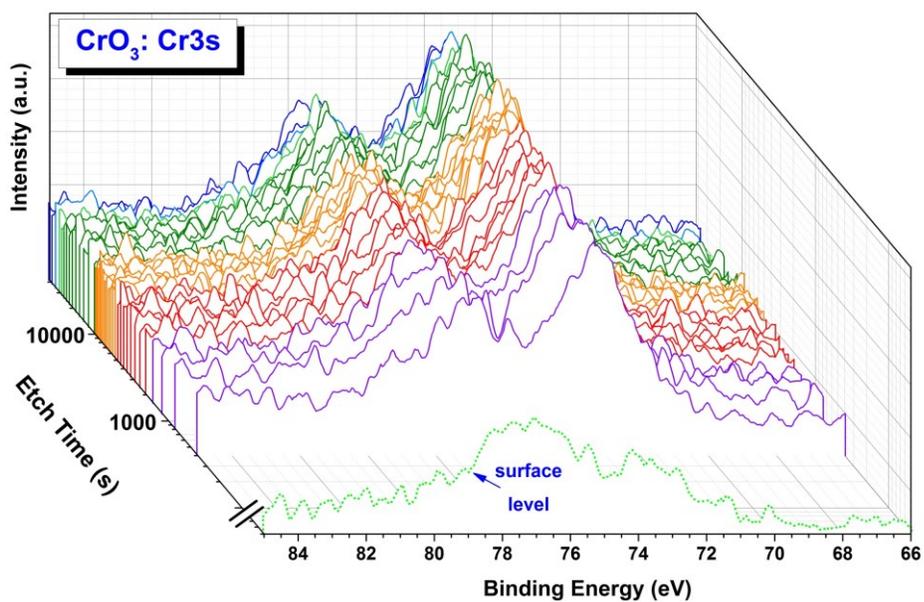

Fig.6. XPS depth profiling of Cr (VI) oxide, with high resolution scans of the Cr2p and Cr3s levels.

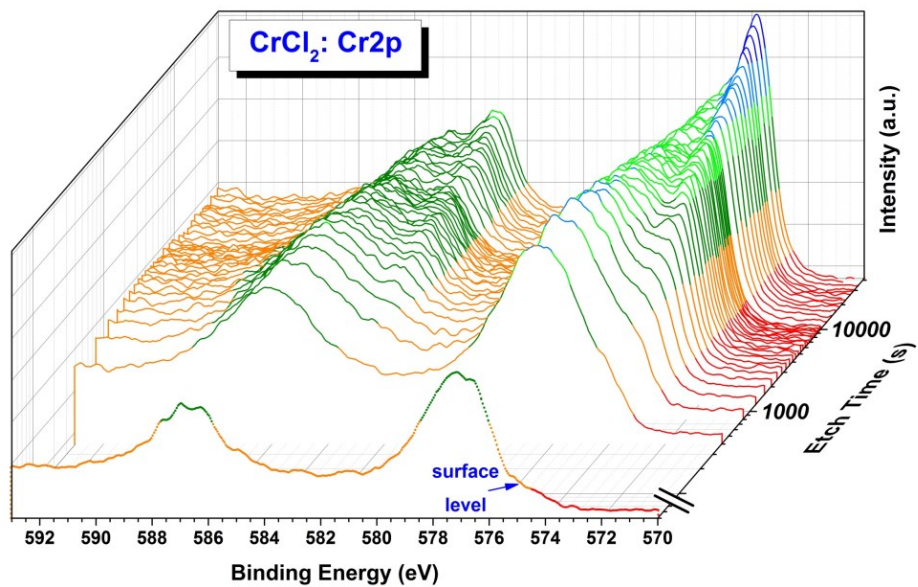

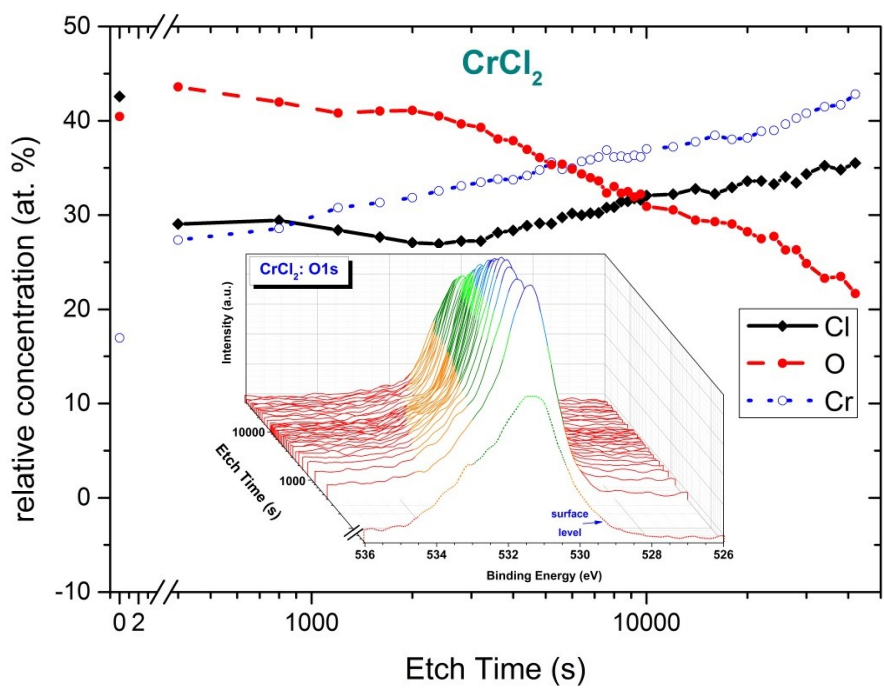

Fig.7. XPS depth profiling of Cr (II) chloride; high resolution scans of Cr2p and the elemental composition (Cr, O, Cl) over time, including spectra of O1s in the inset, are shown.

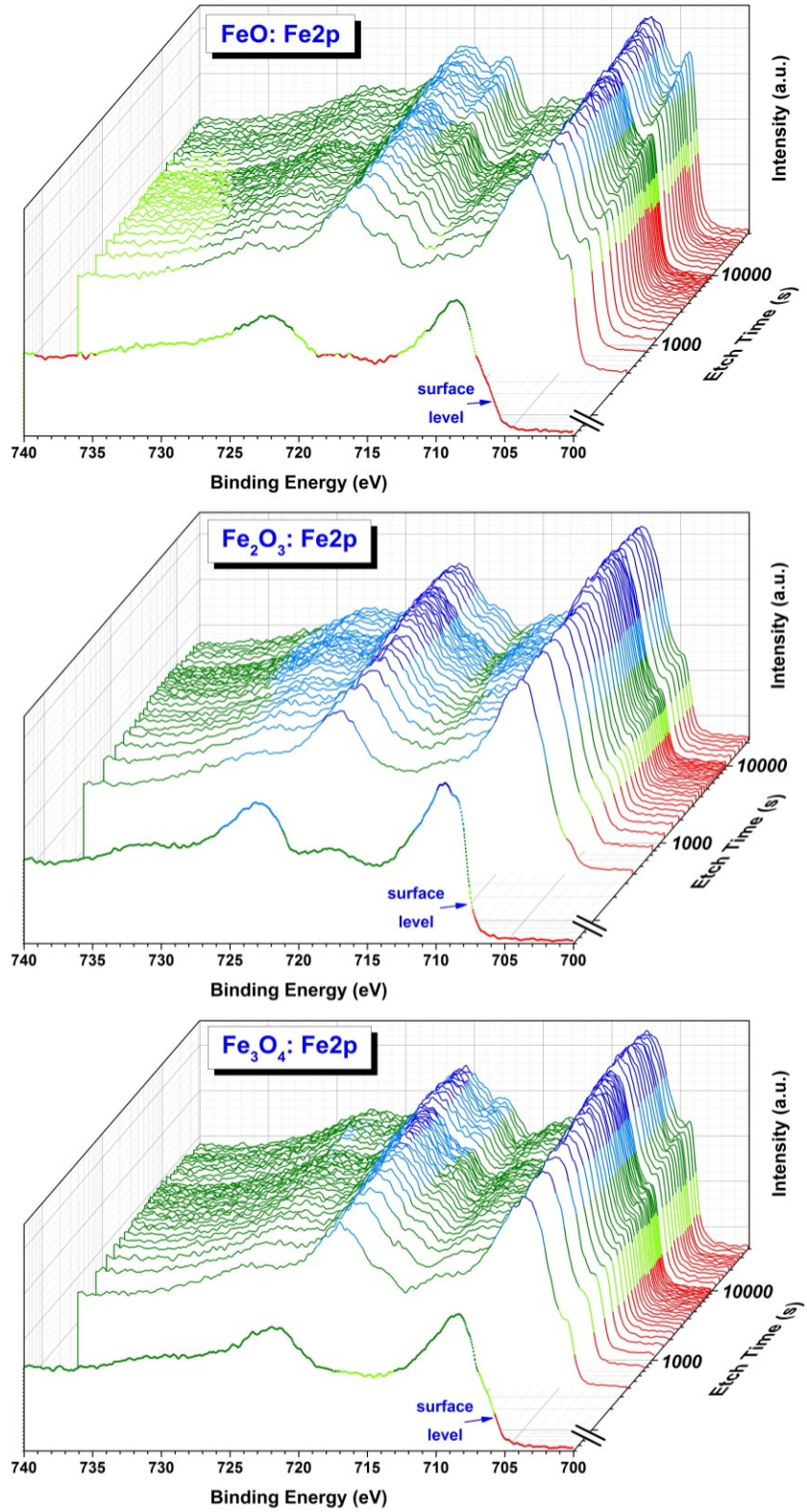

Fig.8. XPS depth profiling of Fe (II), Fe (III) and Fe (II, III) oxide, showing high resolution scans of the Fe2p photo electron level.